\documentclass[conference]{IEEEtran}
\IEEEoverridecommandlockouts
\usepackage{cite}
\usepackage{amsmath,amssymb,amsfonts}
\usepackage{algorithm}
\usepackage{algpseudocode}
\usepackage{graphicx}
\graphicspath{ {./images/} }
\usepackage{textcomp}
\usepackage{xcolor}
\usepackage{fancyhdr}
\usepackage{breakurl}
\usepackage[breaklinks]{hyperref}
\def\BibTeX{{\rm B\kern-.05em{\sc i\kern-.025em b}\kern-.08em
    T\kern-.1667em\lower.7ex\hbox{E}\kern-.125emX}}
    
\begin{document}

\title{Random Linear Network Coding on Programmable Switches\\
}

\author{\IEEEauthorblockN{Diogo Gon\c{c}alves}
\IEEEauthorblockA{\textit{LASIGE, Faculdade de Ci\^encias} \\
\textit{Universidade de Lisboa}\\
Lisbon, Portugal \\
dfgoncalves@fc.ul.pt}\\
\IEEEauthorblockN{Fernando M. V. Ramos}
\IEEEauthorblockA{\textit{LASIGE, Faculdade de Ci\^encias} \\
\textit{Universidade de Lisboa}\\
Lisbon, Portugal \\
fvramos@ciencias.ulisboa.pt}
\and
\IEEEauthorblockN{Salvatore Signorello}
\IEEEauthorblockA{\textit{LASIGE, Faculdade de Ci\^encias} \\
\textit{Universidade de Lisboa}\\
Lisbon, Portugal \\
ssignorello@ciencias.ulisboa.pt}\\
\IEEEauthorblockN{Muriel M\'{e}dard}
\IEEEauthorblockA{\textit{Research Laboratory of Electronics} \\
\textit{Massachusetts Institute of Technology}\\
Cambridge, MA, USA \\
medard@mit.edu}

}

\maketitle
\thispagestyle{fancy}

\begin{abstract}
By extending the traditional store-and-forward mechanism, network coding has the capability to improve a network's throughput, robustness, and security.
Given the fundamentally different packet processing required by this new paradigm and the inflexibility of hardware, existing solutions are based on software.
As a result, they have limited performance and scalability, creating a barrier to its wide-spread adoption.
By leveraging the recent advances in programmable networking hardware, in this paper we propose a random linear network coding data plane written in P4, as a first step towards a production-level platform.
Our solution includes the ability to combine the payload of multiple packets and of executing the required Galois field operations, and shows promise to be practical even under the strict memory and processing constraints of switching hardware.
\end{abstract}


\section{Introduction}
\label{sec:intro}

Network Coding~\cite{ahlswede2000network} is a field in information theory that breaks with the traditional assumption that information relays in networks (e.g., routers) separately carry different information flows.
In Network Coding (NC), to the traditional store-and-forward mechanism we confer the capability to network nodes of \emph{combining} packets.
Specifically, intermediate nodes can mix the information of several packets to generate new packets that are coded combinations of the input.
This principle of operation has been demonstrated to introduce several benefits, including improved throughput~\cite{Katti2008}, robustness against network losses~\cite{Lun2008}, and security~\cite{Lima2007}.

Network coding requires, however, substantially different packet processing from the one available on today's IP-oriented network equipment.
The development of a network coding data plane is faced with non-trivial challenges related to the complex processing logic involved.
First, the need to \emph{combine the payload} of multiple packets.
Traditional packet processing in networking gear is restricted to header manipulation, and payload processing is typically left out of the critical path and offloaded to specialized equipment (consider a DPI middlebox, for instance).  
Second, the \emph{complex Galois field operations} involved in network coding, including finite field multiplication, are challenging to run effectively on a switch, due to its processing and memory constraints.
The requirement of data plane algorithms to process packets at the switch's line rate exacerbate the challenge.

So far, this difficulty was insurmountable: deploying NC on traditional networking hardware was not possible given the incapability to change or extend its operation to meet the requirements of this new paradigm.
Existing implementations are, therefore, based on software, with the inevitable limitations on performance and scalability.
Even high-performance solutions that use specialized hardware, such as FPGA-~\cite{Kim2013} or GPU-based solutions~\cite{Shojania2009}, achieve throughputs that are many orders of magnitude slower than a hardware data plane (e.g., ~\cite{BarefootTofino,BroadcomTomahawk}).
In addition, these solutions target specific systems, making them difficult to port.
We argue that the lack of a network coding data plane with performance and scalability equivalent to its IP counterpart, and the lack of portability of existing solutions to be two of the fundamental barriers for the wide-spread adoption of network coding.

Fortunately, the emergence of production-level programmable switches~\cite{bosshart2013,BarefootTofino,BroadcomTomahawk}, and of a high-level language such as P4~\cite{bosshart2014} to program them, creates exciting new opportunities for architectural innovation in networking.
NC can be seen as a timely example, as per recent proposals for joint efforts between the COmputing In the Network (COIN) and Network Coding (NWCRG) IETF/IRTF research groups~\cite{coin}.
The community now has the opportunity to explore new network architectures and protocols \emph{and} to deploy them in their production networks.
Further, programmability enables architectural evolution \emph{in-situ}, by means of data plane reconfiguration in the field.  
\begin{figure*}[t!]
\centering
\includegraphics[width=1\textwidth,height=0.31\textheight,keepaspectratio]{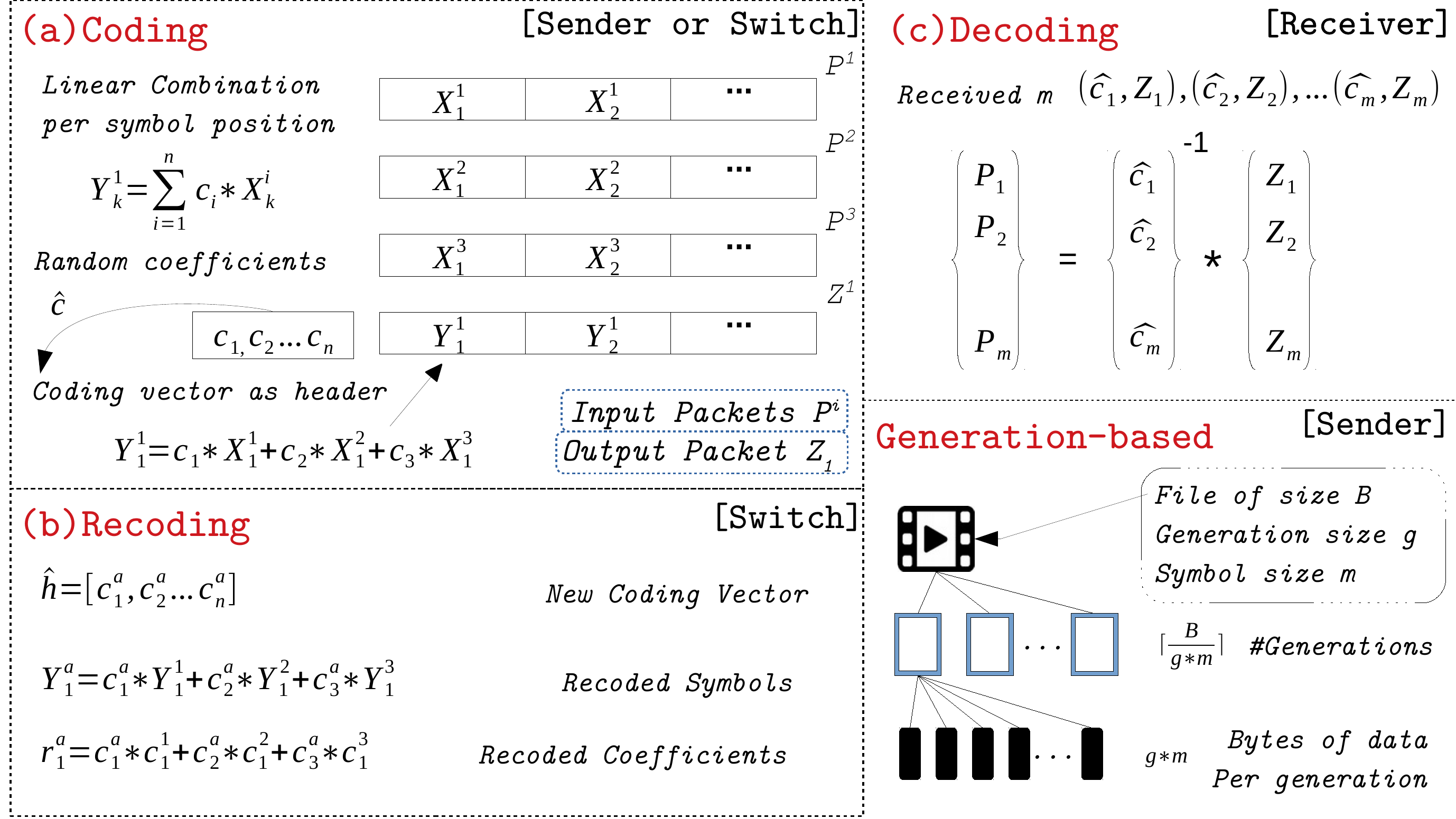}
\caption{Illustration of the encoding (a), recoding (b), and decoding (c) operations in Random Linear Network Coding (RLNC). Each block indicates where (sender/switch/receiver) the respective operation can be performed in our solution. Our data-plane implements a generation-based RLNC, performing coding and decoding operations over generations, that is, blocks of pre-determined size of the transmitted file (see Sec. 2 of \cite{2011code}).}
\label{fig:NCprimer}
\end{figure*}
For our specific purpose, this (r)evolution  makes it possible, for the first time, to build a high-performance, production-quality network coding data plane.
Programmable switching chips can process several billion packets per second, which is orders of magnitude higher throughput than existing NC solutions are capable of.
The use of the P4 language and the increasing number of compilers available also facilitates portability across different software and hardware targets (with minor to no modification to the original programs).

In this paper, we leverage the recent advances in programmable networking hardware and present the design and implementation of a network coding data plane in P4.
Our solution performs random linear network coding~\cite{ho2006random} (RLNC).
The choice for this NC approach is based on its practicality: by decentralizing code generation computation, RLNC avoids the drawback of other NC approaches.
Our preliminary evaluation sheds light on some of the trade-offs involved in a practical data plane implementation of NC.
\section{A Primer on Network Coding}
\label{sec:NCprimer}
Network coding allows a network node to combine several input packets into one coded packet.
In Linear Network Coding, encoding consists of linear operations performed over a finite field.
The process is as follows.

Data carried in packets' payload are interpreted as elements, called symbols, over a finite field $GF(2^m)$ of size $m$.
The \textbf{encoding} process (Figure~\ref{fig:NCprimer}(a)) consists in combining the packets' symbols from incoming packets by using a vector of coefficients $\hat{c}$, called the \textit{encoding vector}, chosen from $GF(2^m)$.
A linear combination is then performed for every symbol position $i$ through the following operation: $Y_k = \sum_{i=1}^{n} c_i * X_k^i$.
The \textbf{decoding} process (Figure~\ref{fig:NCprimer}(c)) allows reconstruction of the original packets from the coded ones.
To recover the transmitted symbols, a receiver node waits until it has enough independent linear combinations (i.e. degrees of freedom) and then performs Gaussian elimination to solve the resulting linear system.
Specifically, assuming a node has a sufficient number of coding vectors and coded symbols $(\hat{c}_m,Y_m)$, it can invert the matrix of the coefficients and resolve the system of linear equations $X=C^{-1} * Y$ to recover the original symbols $X_k$.
Finally, \textbf{recoding} (Figure~\ref{fig:NCprimer}(b)) is the process where coded (or partially decoded) symbols are re-encoded without first being fully decoded.
\subsection{Generation-Based, Random Linear Network Coding}
\label{subsec:genRLNC}
Linear network coding assumes centralized computation of the coding coefficients, limiting its applicability.
However, it is possible to perform network coding in a fully distributed way with \textbf{Random Linear Network Coding} (\textbf{RLNC}) \cite{ho2006random}, by allowing the nodes to choose their linear coefficients independently and uniformly at random over all elements of the finite field.
The encoding vector is then transmitted as an additional header with the coded symbols in the payload.
By choosing coefficients randomly from a sufficiently large finite field it is possible to guarantee coded symbols to be linearly independent with a very high probability ($GF(2^8)$ is usually enough in practice~\cite{2011code}). 
Recoding increases the probability for different nodes in the network to transmit different combinations towards a receiver.
The size of the Galois field and the number of packets to combine affect the computational complexity of the decoding process, the decoding delay and the packet overhead (to transport the coding vectors).
A practical technique to bound these effects consists in grouping packets into blocks, called \textbf{generations}, over which coding and decoding are performed.
The \emph{generation size} defines the number of symbols over which coding is performed.
A large generation size improves throughput by increasing the probability of generating independent combinations, while increasing decoding complexity and network overheads.
Generally, the optimal trade-off between throughput and complexity/overhead depends on the network topology and on the specific application \cite{2011code}.
\section{Network Coding in the data plane}
\label{sec:ncDataPlane}
The design of a network coding solution requires i) a protocol for end-hosts and relay nodes in the network to carry and exchange coding parameters, and ii) specific modules in the switches to implement the coding operations.
With regard to the former, we define a packet format enabling the implementation of a generation-based RLNC protocol exchange.
With respect to the latter, we design software modules for the data-plane of a P4-programmable switch implementing the coding and recoding functions.
The design of the switch's modules copes with the limited expressiveness of the P4 language \cite{p4spec16}.

\subsection{Packet Format} 
The packet format used by our generation-based RLNC protocol contains two headers, an inner and an outer header, carried over Ethernet frames.
The inner header carries a wire representation of the symbols according to the encoding proposed in \cite{I-D.heide-nwcrg-rlnc}.
This header contains symbols (and coding vectors when present) prepended, among others, by information about the number of symbols encoded in the packet and the type of packet (either coded or uncoded).
The outer header contains coding parameters set by  an exchange between sender and receiver nodes,  according to the guidelines in \cite{I-D.heide-nwcrg-rlnc-background}, which are then used by switches for NC operation.
The proposed outer header specifies the generation identifier, and the generation, finite field and symbol sizes.\\
\subsection{Buffering and Packet Replication}
In real networks, packets of the same generation are not carried synchronously over the same links.
They experience different propagation and queuing delays, can traverse different paths, and are thus received by each node at different times, and in potentially different order.
Therefore, symbols and coding vectors from a certain generation need to be stored by a switch until a sufficient number of symbols is received, to trigger coding of packet(s) from that generation.

In our design, this is achieved by a buffering module shared across different generations, and indexed by the generation identifier.
By limiting the maximum number of generations concurrently stored and the generation size, we bound the memory resources required by this component.
Moreover, a generation is flushed from a switch's buffer upon the reception of an acknowledgement issued by a receiver to signal the successful decoding of that generation.

When this buffer contains a number of symbols that is greater or equal to the generation size, linear combinations of the symbols of the current generation can be produced by the switch.
The reception of the last packet of a generation thus triggers the creation of one or more linearly coded packets from the current generation.
The process produces coded symbols (and coefficients, if recoding is performed) which are inserted into the inner header of one or more coded packets.
The outer header is kept the same as the original input packets.

In our design, the number of linear combination to be generated and transmitted is a parameter which can be reconfigured at run-time, in the field, through the control plane.
This flexibility is important, as the number of packets required for a receiver to successfully decode a generation may vary with network conditions. 

\begin{algorithm}
\caption{Russian Peasant Multiplication}
\label{alg:russianMul}
\hspace*{\algorithmicindent} \textbf{Input} ${\alpha,\beta} \in GF(2^m)$, $s$ is bit-length of the factors $\alpha$ and $\beta$, $\delta$ is an irreducible polynomial for $GF(2^m)$
\begin{algorithmic}[1]
\Procedure{MultiplyGFelements}{$\alpha,\beta$}
\State $product \gets 0$
\For{$i=0$ to $m-1$} \Comment{unrolled in m action calls}
\State $product \gets product \oplus (-(\beta \land 1) \land \alpha)$
\State $mask \gets ((\alpha >> (m-1)) \land 1)$
\State $\alpha \gets (\alpha << 1) \oplus (\delta \land mask)$
\State $\beta \gets \beta >> 1$
\EndFor
\State return $product$
\EndProcedure
\end{algorithmic}
\end{algorithm}
\subsection{Finite Field Arithmetic}
Computing a linear combination of the symbols of a generation requires i) selecting uniformly at random coefficients in the finite field, ii) multiplying the symbols by the selected coefficients, and iii) summing the resulting factors (Section~\ref{sec:NCprimer}).
Hence, RLNC requires the generation of random numbers and arithmetic operations to be performed in a Galois field $GF(2^m)$.

The selection of coefficients can be achieved by leveraging hardware random number generators, which are widely present across commercial switching chips.
Finite field addition is a regular polynomial addition that can be performed by standard bitwise-xor.
Multiplication in $GF(2^m)$ is, however, more challenging to implement, since it requires reducing, through a modulo operation, the product of the two factors by an irreducible polynomial over the finite field.\\
Due to its relevance for several applications, including cryptography and error detection/correction, efficient software techniques \cite{ning2001efficient} for finite field multiplication have been proposed.
In our finite field arithmetic module we feature two multiplication techniques with different characteristics: one compute-intensive, and one memory-intensive.
The first is a Standard Field Multiplication algorithm \cite{kul15}, shown in Alg. \ref{alg:russianMul}, which can be implemented with simple shift and add operations.
This is an iterative algorithm which operates on the two factors bit by bit.
The second technique is based on pre-computed lookup tables.
It exploits the property that every element $x \neq 0 \in GF(2^m)$ can be represented uniquely by power to a primitive element $\delta$, so $x \equiv \delta^i$, where $i$ is the discrete logarithm of $x$ with respect to $\delta$.
Because of this property, the multiplication of two elements ${\alpha,\beta} \in GF(2^m)$ can be rewritten like $mul(\alpha,\beta) \equiv antilog((log(\alpha) + log(\beta)) mod Q)$ where $Q$ is the size of the finite field.
The log/antilog values can be pre-computed based on the primitive element $\delta$, and be stored in tables for lookup at packet processing time (e.g., $512 B$ are required to store these tables for the case under analysis $GF(2^8)$).
With this technique, a multiplication in $GF(2^m)$ involves 3 table look-ups, 1 addition and 1 modulo operation.
As the modulo operation is computationally expensive and may not be supported by some targets, we avoid it by employing an optimization based on~\cite{greenan2008optimizing}.
The end result is presented in Alg. \ref{alg:logMul}.

\begin{algorithm}
\caption{Log/Antilog Tables Multiplication}
\label{alg:logMul}
\hspace*{\algorithmicindent} \textbf{Input} ${\alpha,\beta} \in GF(2^m)$, $Q \equiv 2^m$ is the field size, log/antilog values are stored in registers 
\begin{algorithmic}[1]
\Procedure{MultiplyGFelements}{$\alpha,\beta$}
\If{$\alpha == 0 || \beta == 0$}
    \State return 0
\EndIf
\State $sum \gets log[\alpha] + log[\beta]$
\If{$sum \geq Q - 1 $}
\State $sum \gets sum - (Q - 1)$    
\EndIf

\State return $antilog[sum]$
\EndProcedure
\end{algorithmic}
\end{algorithm}

\section{The RLNC.P4 program}
\label{sec:p4Program}
\begin{figure}[t]
\centering
\includegraphics[width=0.5\textwidth,keepaspectratio]{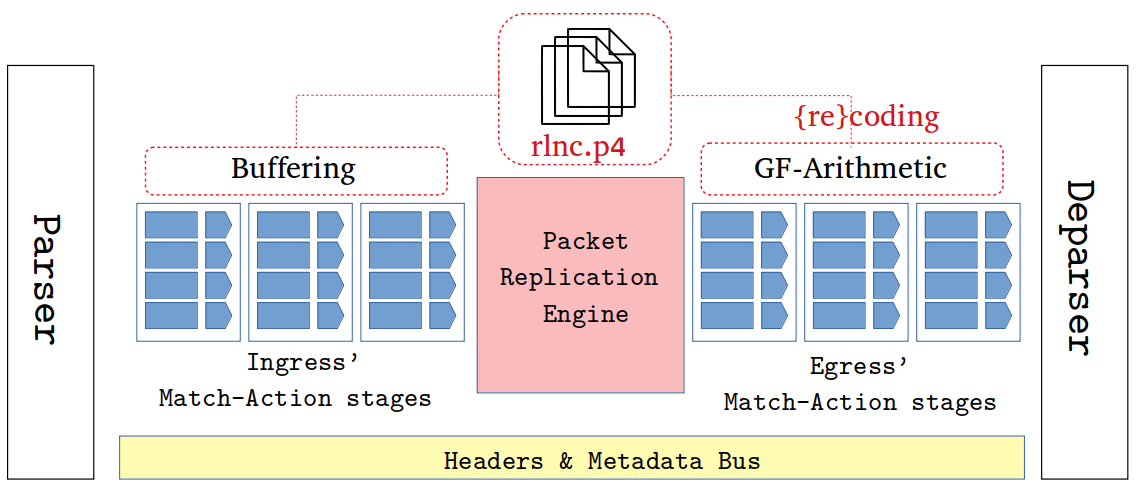}
\caption{Packet processing of our network coding switch on a PISA-like architecture.}
\label{fig:programOnPisa}
\end{figure}

This section presents the main modules of our network coding switch, implemented in P4-16\footnote{Available open-source at https://github.com/netx-ulx/NC.}.
Figure \ref{fig:programOnPisa} illustrates how the different modules are mapped to the processing pipelines of a PISA-like switch architecture.
Packets are first buffered in the ingress pipeline.
For each incoming packet the (coded or uncoded) symbols in the packet payload, as well as the coefficients (only carried with coded and recoded symbols), are buffered into the switch's stateful memories.
If the buffer that maintains that packet's generation is not yet filled, i.e., if it does not contain enough symbols to start coding the current generation, the packet is dropped.
Otherwise, the ingress pipeline sets the necessary metadata for the Packet Replication Engine (PRE) to produce the necessary packet copies (the exact number is configured by the control plane), through the use of a multicast primitive.

Each packet replica goes through the egress pipeline where different coefficients are selected through a random number generator primitive.
Then, the egress pipeline executes the arithmetic module to create the coded symbols.
The coding process is different whether or not the buffered symbols are coded or uncoded.
As illustrated in Fig. \ref{fig:NCprimer}, while encoding involves computing only linear combinations of the symbols, recoding linearly combines also the encoding vector.
Two different modules are available to perform the necessary arithmetic over finite fields, implementing the two multiplication algorithms presented in Sec. \ref{sec:ncDataPlane}.
This module can be configured at code generation time.
Packets carrying linear combinations for a generation are generated and forwarded by the switch until an acknowledgement packet is received.
This acknowledgement is generated by the receiver once it has received a sufficient number of linearly independent symbols that allows decoding that generation.
Once the switch receives an acknowledgment packet, it stops producing linear combinations for this generation, and frees up the corresponding buffer space.
\subsection{Discussion and lessons learned}
\label{subsec:practicalConsideration}
Our program fully captures the targeted coding behavior.
Yet, there are target-specific factors that may limit its applicability and/or affect its functionality.
We discuss these here, with the aim to potentially identify design patterns useful to the evolution of the language and/or the related targets.

\textbf{Target-specific.}
\emph{Parsing Payload.} The exact number of symbols and coefficients in each packet is unknown at compile-time.
These fields are extracted by the program into P4 headers at run-time, in order to be buffered in the ingress pipeline.
Moreover, although the maximum number of symbols in each packet is defined by the packet format, the length of the coding vector for each encoded symbol varies with the generation size.
This typically results in a large header vector, which contrasts with the limited size of the headers' bus on P4-programmable targets, that is usually much smaller than the typical packet payload  (e.g., a few kbits in PISA architectures~\cite{bosshart2013}).
For protocols with small payloads, it is possible to treat the symbols as headers, as in our solution.
However, when the payload is large, either the target needs to have a wider header vector bus, or it becomes necessary to pre-process packets to split them into smaller fragments.
This is something we plan to investigate as future work.
Any potential application must therefore consider packet sizes accordingly to the target's ability to extract the necessary information into the header vector.

%
\emph{Packet Replication.} Several packet copies may be generated by the PRE and edited by the egress pipeline to encode a buffered generation.
To avoid the throughput penalty of recirculating packets, our program leverages the multicast primitive available in several targets for this purpose.
One issue in some targets is that the so-generated packet copies may need to be forwarded through different output ports.

\textbf{Language-specific.} 
We developed a code-generating template to tailor the program with respect to the various coding parameters, namely the maximum number of concurrently supported generations, the generation size, etc.
This dependency on coding parameters primarily affects the size of the program, yet it has also influenced its software design.

\emph{Buffering} is implemented through target's externs for stateful memory, namely registers, which are supported by several P4-programmable targets.
This requires memory indexed by generation identifiers to be dynamically-allocated for storing symbols and coefficients.
However, registers' length and cell size must be specified at compile-time.
Besides, read/write operations on registers are methods invoked on the register itself rather than calls to a generic method with a register's name as parameter.
Given these constraints, we implemented buffering with a single register partitioned among generations.
This approach enables a dynamic allocation of the buffer space to different generations (and results in a less verbose and more modular program).
Yet, this design choice requires complementary registers to store base and offset pointers per generation.
\textit{This poses the question of whether or not P4-programmable architectures should expose anything more specific than a general purpose register.}

\emph{Coding and Recoding} operations depend on the generation size, which defines the number of operands in each linear combination to be computed.
Every multiplication and addition requires similar action calls.
The code-generating template we developed automates this process, but this has an impact on code readability and re-usability.
\textit{Would a set of primitives or annotations for loops, e.g., to be unrolled by a preprocessor \cite{shah2018pcube}, produce a less error-prone and easier-to-read code?}
\begin{figure}[t]
\centering
\includegraphics[scale=0.35,keepaspectratio]{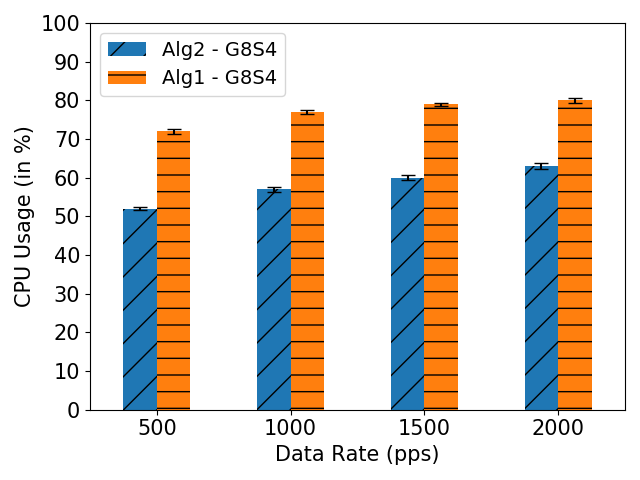}
\caption{CPU usage (in \%) for the two algorithms with generation size 8 and 4 1-byte symbols per packet.}
\label{fig:cpuUsage}
\end{figure}
\section{Evaluation}
\label{sec:evaluation}
We ran a preliminary evaluation of our RLNC P4 program on the reference P4 software switch \cite{bmv2}.
The experiments were run on a bare-metal Dell PowerEdge R440 server with 2x Intel Xeon Silver 4114, 2.2GHz, and 32GB of memory.
For all the reported results, ten independent experiments have been executed and average values are plotted.
Through this first evaluation, we have tried to assess
i) differences in the performance between the two multiplication algorithms we implemented, 
ii) the impact of coding parameters on network throughput, and
iii) the throughput penalty of performing recoding.

\subsection{Program Size} 
Coding parameters, including the generation size and the number of symbols in the coded packets, affect the size of the code in the buffering and arithmetic modules of the program.
Therefore, we have explored different configurations of these parameters and measured the size of the corresponding P4 programs.
Both multiplication algorithms, presented in Sec. \ref{sec:ncDataPlane}, have been tested, as they have different requirements (Alg. \ref{alg:russianMul} is compute-intensive whereas Alg. \ref{alg:logMul} is memory-intensive). Overall, Alg. \ref{alg:russianMul} produces more compact programs. Yet, we have found the compiled version of programs featuring Alg. \ref{alg:russianMul} to be up to 3x orders of magnitude larger than the Alg. \ref{alg:logMul} counterpart. We are still investigating the root cause of this difference. We are trying to understand if a different programming style would reduce it, and are also looking deeper at the target compiler. One possibility is the fact that this algorithm performs bit-by-bit operations that may be a poor fit to the target architecture. Given this issue and the fact that this algorithm consumes 20\% higher CPU (see Fig. \ref{fig:cpuUsage}), we have focused the rest of our performance analysis on Alg. \ref{alg:logMul}.\\
\begin{figure}[t]
\centering
\includegraphics[scale=0.35,keepaspectratio]{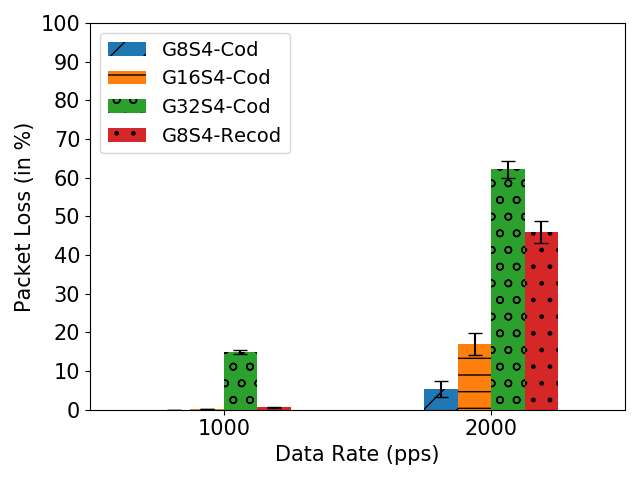}
\caption{Packet Loss in (\%) for coding (-cod) and recording (-recod) with different configurations of generation size Gx and symbols per packet Sy.}
\label{fig:pktL}
\end{figure}
\subsection{NC Switch Performance}
Across all the tested coding configurations, we stressed the RLNC program to data-rates that start introducing packet loss.
In particular, Fig. \ref{fig:pktL} illustrates that either dealing with a larger generation size (e.g., 16 or 32) or recoding start generating high packet loss rates.
This was expected since an increase $x$ of the generation size determines an increase of the coding vector per each symbol and, by consequence, an increase of a factor of $x*n$ on the number of required arithmetic operations, where $n$ is the number of symbols per packet.
Recoding also suffers from larger generation sizes, since it requires parsing and storing more elements and performing additional arithmetic over the coding vectors.\\
Overall, these preliminary results show that generation size and recoding affect the performance of our switch's data-plane.
In practice, however, large generation sizes are not common across network coding applications, as they increase the computation complexity of the decoding process and introduce large packet overheads, as we also observed.
Furthermore, the impact of both factors can be considerably reduced by leveraging both sparse coding techniques and coding over overlapping generations, optimizations which we aim to investigate in our future work.
\section{Final remarks}
\label{sec:conclusion}
The networking research community has for a long time struggled with the ossification of the Internet~\cite{Peterson2003}, a result of the interplay of its original design and the vested interests of competing stakeholders.
Addressing its various problems was limited to incremental changes, stifling innovation and precluding disruptive architectural advances.
Several research projects proposed clean-slate redesigns of the Internet architecture, but they have been restricted to software-based implementations and small research testbeds.
As a result, so far these radical designs have all shared the same fate: as there was no clear way to migrate from the research testbed to a large-scale, high-performance production network, they have not left the research lab.
We believe programmable ASICs and P4 to allow, for the first time, the implementation of radical architectural approaches in high-speed hardware, improving their prospect for Internet-wide deployment (e.g., through backwards-compatible frameworks such as Trotsky~\cite{McCauley2019}).
We made a first attempt in~\cite{Signorello2016,Miguel2018}, targeting NDN.
The work we presented here is inserted in this same line but targets a different paradigm -- network coding -- with fundamentally diverse challenges.

\section*{Acknowledgment}
We would like to thank the anonymous reviewers for their feedback, which helped improve the paper.
This work was supported by FCT through funding of the uPVN project, ref. PTDC/CCI-INF/30340/2017, and LASIGE Research Unit, ref. UID/CEC/00408/2019.
\bibliographystyle{IEEEtran}
\bibliography{sample-base}

\end{document}